\PassOptionsToPackage{shortlabels}{enumitem}
\documentclass[11pt, a4paper]{include/gdm_format}

\usepackage[authoryear, sort&compress, round]{natbib}
\usepackage{xcolor}
\usepackage{subcaption}
\usepackage{mwe}
\usepackage{graphicx}
\usepackage{xr}
\usepackage{marvosym}

\usepackage[frozencache,cachedir=.]{minted}

\usepackage{fvextra}
\usepackage{xcolor}

\definecolor{bgcolor}{rgb}{0.95,0.95,0.95}

\setminted{
    bgcolor=bgcolor,       %
    fontsize=\footnotesize,
    breaklines,
}

\usepackage{tablefootnote}
\usepackage{pdfpages}
\usepackage{minitoc}
\usepackage{titletoc}
\usepackage{appendix}

\usepackage{hyperref}
\usepackage{url}
\usepackage{xurl}
\usepackage[parfill]{parskip}

\usepackage{amsmath}
\usepackage{xspace}
\usepackage{multirow}
\usepackage{booktabs}
\usepackage{color}
\usepackage{colortbl}
\usepackage{cleveref}
\usepackage{graphicx}
\usepackage{algorithm}
\usepackage{algorithmicx}

\usepackage{algpseudocode}
\usepackage{subcaption}
\usepackage{wrapfig}
\usepackage{sidecap}
\usepackage{soul}
\sidecaptionvpos{figure}{t}
\usepackage{multicol}
\usepackage{pifont}
\usepackage[normalem]{ulem}
\usepackage{listings}
\usepackage[most]{tcolorbox}

\title{ReXGradient-160K: A Large-Scale Publicly Available Dataset of Chest Radiographs with Free-text Reports}

\correspondingauthor{Pranav Rajpurkar, pranav\_rajpurkar@hms.harvard.edu}

\author[1]{Xiaoman Zhang}
\author[1]{Juli\'an N. Acosta}
\author[2]{Josh Miller}
\author[2,3,4]{Ouwen Huang}
\author[1]{Pranav Rajpurkar \Letter}

\affil[1]{Department of Biomedical Informatics, Harvard Medical School, Boston, MA, USA}

\affil[2]{Gradient Health, Durham, NC, USA} 

\affil[3]{Department of Statistical Science, Duke University, Durham, NC, USA}

\affil[4]{Laplace Institute, Durham, NC, USA}

\begin{document}

\begin{abstract}
We present ReXGradient-160K, representing the largest publicly available chest X-ray dataset to date in terms of the number of patients. 
This dataset contains 160,000 chest X-ray studies with paired radiological reports from 109,487 unique patients across 3 U.S. health systems (79 medical sites).
This comprehensive dataset includes multiple images per study and detailed radiology reports, making it particularly valuable for the development and evaluation of AI systems for medical imaging and automated report generation models. The dataset is divided into training (140,000 studies), validation (10,000 studies), and public test (10,000 studies) sets, with an additional private test set (10,000 studies) reserved for model evaluation on the \href{https://rexrank.ai}{ReXrank} benchmark. By providing this extensive dataset, we aim to accelerate research in medical imaging AI and advance the state-of-the-art in automated radiological analysis.
Our dataset will be open-sourced at \url{https://huggingface.co/datasets/rajpurkarlab/ReXGradient-160K}.
\end{abstract}

\maketitle

\section{Introduction}
The increasing global demand for radiological expertise, coupled with uneven distribution of specialists and growing workloads, has created significant challenges in healthcare delivery.
Recent years have seen remarkable advances in artificial intelligence (AI) applications for medical imaging, particularly in developing AI systems that can generate comprehensive radiology reports, aiming to enhance workflow efficiency and expand access to expert-level interpretations~\citep{tanida2023interactive,zhou2024generalist,bannur2024maira,pellegrini2023radialog}.

Several large-scale chest X-ray datasets with paired reports have been instrumental in advancing this field. The MIMIC-CXR dataset, comprising 227,835 studies from 65,379 patients treated at Beth Israel Deaconess Medical Center, provided one of the first large-scale collections of radiographs with corresponding free-text reports~\citep{johnson2019mimic}. 
This was followed by CheXpert Plus, which enhanced the original CheXpert dataset with detailed radiology reports and metadata across 223,228 studies from 64,725 patients, offering improved capabilities for model development and evaluation~\citep{chambon2024chexpert}.
The IU X-ray dataset, a smaller database with 7,470 image-report pairs, has been widely used for early development and testing of report generation models ~\citep{demner2016preparing}.

However, as AI-assisted medical reporting continues to evolve rapidly, there is a critical need for standardized benchmarks and comprehensive evaluation frameworks. Existing datasets often face limitations regarding consistent data splits, standardized evaluation metrics, and the ability to test model generalization across different clinical settings~\citep{johnson2019mimic,chambon2024chexpert}. Most notably, these datasets are typically collected from single institutions, which limits their ability to evaluate models' generalization capabilities across diverse healthcare settings.
The Medical AI Data for All (MAIDA) initiative~\citep{saenz2024maida} represents another important effort to address the generalizability challenge by establishing a framework for global medical-imaging data sharing across diverse clinical environments. However, its current data collection scale of approximately 100 scans per setting limits its ultimate size.

To address these challenges, we present ReXGradient-160K, a large-scale, multi-institutional dataset of chest radiographs with paired radiology reports. Our dataset contains 160,000 chest X-ray studies with associated reports from over 100,000 unique patients across multiple medical institutions, making it the largest publicly available multi-site chest X-ray dataset to date. This dataset is complemented by ReXGradient, a private evaluation set of 10,000 studies from 67 U.S. medical sites, which serves as the test set for the ReXrank benchmark (\url{https://rexrank.ai})~\citep{zhang2024rexrank}. Together, this comprehensive collection includes multiple images per study and detailed radiology reports, making it particularly valuable for developing and evaluating AI systems for medical imaging and automated report generation. The multi-institutional nature of our dataset provides a unique opportunity to assess and improve the generalization capabilities of AI models across different clinical settings and geographical locations.

\section{Dataset Composition}
ReXGradient-160K comprises 273,004 unique chest X-ray images from 160,000 radiological studies, collected from 109,487 unique patients across 3 U.S. health systems. The dataset is divided into three splits: training (140,000 studies), validation (10,000 studies), and public test (10,000 studies) sets. A single patient may be associated with multiple studies over time, and each study typically contains one or more images. Table~\ref{tab:dataset_stats} provides a comprehensive overview of the dataset statistics across different splits, including the number of studies, unique images, patients, and the mean token counts for each report section.


\begin{table}[t]
\centering
\caption{Dataset Statistics. Study Statistics show the distribution of studies, unique images, and distinct patients. Report Statistics display the mean number of tokens for each report section (Indication, Comparison, Findings, and Impression).}
\begin{tabular}{lccccccc}
\toprule
\multirow{2}{*}{\textbf{Split}} & \multicolumn{3}{c}{\textbf{Study Statistics}} & \multicolumn{4}{c}{\textbf{Report Statistics (Mean Tokens)}} \\
\cmidrule(lr){2-4} \cmidrule(lr){5-8}
& Studies & Images & Patients & Indication & Comparison & Findings & Impression \\
\midrule
Train & 140,000 & 238,968 & 95,716 & 5.11 & 2.65 & 32.27 & 11.17 \\
Validation & 10,000 & 17,007 & 6,964 & 5.16 & 2.75 & 32.92 & 11.63 \\
Test & 10,000 & 17,029 & 6,807 & 5.13 & 2.78 & 32.48 & 11.52 \\
\bottomrule
\end{tabular}
\label{tab:dataset_stats}
\end{table}

\subsection{Data De-identification}
Our de-identification process is fully HIPAA compliant and is broken down into 2 parts: text de-identification for reports and imagery metadata, and pixel de-identification for removing protected health information (PHI) found in the imagery specifically.
In the cases of names, IDs and dates we resort to pseudonymization to add an additional layer of protection. Names are replaced, IDs are altered and dates are shifted to within 365 days of the actual date. Studies for the same patient undergo the same date shift in order to maintain validity in comparing patient outcomes over time

\subsection{Image Characteristics}
All images in the dataset were originally in DICOM format and have been converted to PNG format using a standardized preprocessing pipeline. The process began with pixel data extraction from DICOM files using pydicom, followed by proper handling of bit depth and photometric interpretation. For MONOCHROME1 images, pixel values were inverted to maintain consistent representation across the dataset.  Standard min-max normalization to the full 16-bit range (0-65535) was applied. In accordance with our data use agreement, images were downsampled to 25\% of their original dimensions using cubic interpolation with anti-aliasing to maintain important structural details.

\subsection{Report Structure}
Each radiological report in our dataset is structured into four key sections:
\begin{itemize}
\item \textbf{Indication}: Provides relevant patient background and reason for examination
\item \textbf{Comparison}: Provides relevant patient background and reason for examination
\item \textbf{Findings}: Detailed radiological observations
\item \textbf{Impression}: Summary of key conclusions and recommendations
\end{itemize}
To ensure consistent report formatting across all studies, we prompt GPT-4o to extract four key sections from the original reports.  Additionally, we implemented a robust post-processing validation step to ensure all sections contained valid content, remove the cases where particular sections might be missing in the original reports.

\subsection{Demographic Distribution}
Table~\ref{tab:demographics_stats} presents the demographic distributions across training, validation, and testing splits. The age distribution shows consistency across all splits, with patients aged 40-80 years representing the largest proportion (approximately 50\% of the dataset). The sex distribution exhibits a balanced ratio between male and female patients (approximately 49\% male and 50\% female) across all splits, with a minimal percentage (1.3-1.8\%) of cases with unknown sex information. 

\begin{table}[t]
\centering
\caption{Demographic Statistics. Age Statistics show the percentage distribution across age ranges. Sex Statistics display the percentage distribution by gender.}
\begin{tabular}{lccccccccccc}
\toprule
\multirow{2}{*}{\textbf{Split}} & \multicolumn{5}{c}{\textbf{Age Statistics (\%)}} & \multicolumn{3}{c}{\textbf{Sex Statistics (\%)}} \\
\cmidrule(lr){2-6} \cmidrule(lr){7-9}
& 0-20 & 20-40 & 40-60 & 60-80 & 80+ & M & F & U\\
\midrule
Train & 16.9 & 19.4 & 25.1 & 26.0 & 12.6 & 49.0 & 49.5 & 1.5 \\
Validation & 17.9 & 20.3 & 25.5 & 24.0 & 12.2 & 48.3 & 50.4 & 1.3\\
Test & 17.3 & 19.6 & 25.4 & 24.5 & 13.2 & 47.9 & 50.3 & 1.8 \\
\bottomrule
\end{tabular}
\label{tab:demographics_stats}
\end{table}

\section{Data Records}
All data are available on Huggingface. Access to the ReXGradient-160K dataset requires signing our license on Huggingface.
The dataset organizes images into subfolders, with each subfolder named according to the anonymous patient identifier and study ID. 
Each patient subfolder contains one or multiple studies, with each study containing one or multiple images. 
For the reports, we provide metadata including PatientID, AccessionNumber, PatientSex, EthnicGroup, PatientAge, PatientWeight, StudyDate, InstitutionName, and Manufacturer.

\section*{Acknowledgement}
This work was supported by the Biswas Family Foundation’s Transformative Computational Biology Grant in Collaboration with the Milken Institute.

\section*{Disclosures}
O.H and J.M are founders and hold equity in Gradient Health, a private company focused on health data accessibility and availability for commercial research. Gradient Health provided the dataset used in this work and did not provide funding for this research and had no role in its design, execution, or publication. 


\bibliography{references}
\bibliographystyle{plainnat}

\end{document}